\documentstyle[aps,multicol,epsfig]{revtex}
\def\vec#1{{\rm\bf #1}}

\draft

\input epsf

\begin{document}

\title{Theory of Chiral Imprinting}
\author{Y.~Mao, M.~Warner}
\address{
Cavendish Laboratory, Madingley Road,\\
Cambridge, CB3 0HE,  UK.}
\date{\today}
\maketitle


\begin{abstract}
We present a continuum model for a nematic elastomer network
formed in a chiral environment, for instance in the presence of a
chiral solvent. When this environment is removed, the network can
retain some memory of its chiral genesis. We predict the residual
chiral order parameter for a number of possible scenarios, and go
on to examine the robustness (stability) of the imprinted chirality. 
We show that a twist-untwist transition can take place,
which determines whether the imprinting has been successful. 
A transition is via a coarsening of the helical director pattern and
a lengthening of its pitch. Finally, the effect due to a
subsequent swelling by an achiral solvent, or by a solvent of
differing chirality, is considered.
\end{abstract}



\vspace{0.25in} \noindent {PACS numbers:} 61.30.-v, 61.41.+e,
78.20.Ek

\begin{multicols}{2}

 Nematic elastomers combine the properties of a liquid
crystal and those of a conventional rubber. This synergy gives
rise to novel material behaviour, which in turn has stimulated
much research in past years \cite{mark}. From a general symmetry
argument, de Gennes \cite{degennes} first suggested that chirality
may be introduced to such an elastomer by simply forming it in a
chiral solvent. The originally achiral liquid crystalline polymer
would then remember the induced chirality after crosslinking, even
when the solvent is removed or replaced with an achiral one. This
is the case of chiral imprinting, which potentially can open up an
entirely new way of producing materials of specified optical
properties. Chiral imprinting, in principle, is akin to
crosslinking a nematic polymer under an external magnetic or
mechanical field \cite{mark,finklemann} where the monodomain state
is also permanently imprinted. Experimentally chiral imprinting
was studied long ago \cite{Tsutsui} as a function of solvent
exchange. Recently imprinting has been studied \cite{geof} as a
function of both solvent removal and temperature. Another measure
of imprinting occurs in intrinsically cholesteric networks. On
temperature changes that would cause a substantial pitch variation
in a non-crosslinked cholesteric polymer melt, the corresponding
network suffers essentially no variation - see for example Fig.\ 8
of reference \cite{max}, where also many other references to
cholesteric elastomers are given. In this paper, we analyse chiral
imprinting, and predict the retained chiral properties of the
elastomer when the initial chiral environment of crosslinking is
altered. Gradients of director variation are modeled within
continuum Frank nematic elasticity. The nematic elastomer penalty
for rotation of the director relative to the solid matrix is
described in a fully non-linear, rubber-elastic manner since
rotation can be large.


A nematic liquid crystal has a mobile director $\vec{n}$, the
gradient of which incurs a Frank energy \cite{degennesbook}. For
the free energy density of a cholesteric liquid crystal, the twist
term is modified by a pitch wave number, $q_0$: $$
f_c \!= \!\textstyle{1 \over 2}  \!\left[K_1 (\bbox{\nabla} \!\cdot \!
\vec{n})^2 \!+\! K_2 (\vec{n}\! \cdot \!(\bbox{\nabla} \!\times\!
\vec{n})\!+\!q_0)^2 \!+\! K_3 ((\vec{n} \!\cdot \! \bbox{\nabla}) \;
\vec{n})^2 \right] $$ where $K_{1,2,3}$ respectively measure the
energy penalty for splay, twist and bend, the three possible modes
of the nematic director distortion.
 In a pure twisting, cholesteric
conformation of the director, we can drop the terms involving
$K_{1,3}$. These other contributions arise if the director moves
away from the helical plane and tilts toward the pitch axis. The
Frank energy can be viewed as a continuum description with higher
order spatial derivatives truncated, and it suffices for cases
where director varies slowly over a nematic coherence length
($\sim 10$nm).

When liquid crystalline polymers are crosslinked into a rubber
network, additional constraints on $\vec{n}$ arise in the form of
director anchoring to the network. Anchoring manifests itself with
an extra energy cost \cite{degennes2} for a uniform director
rotation $\bbox{\omega}$ relative to a local rotation of the
elastic  matrix $\bbox{\Omega}$. The energy density is
$\textstyle{1\over 2} D_1 \; [(\bbox{\Omega}-\bbox{\omega})\times
\vec{n}]^2
$. A second term couples the relative director-matrix rotation to
the shearing part of the elastic strain, $\bbox{{\lambda}}$, that
is $D_2 \;\;\vec{n}\; .\; \bbox{{\lambda}}\;
.\;[(\bbox{\Omega}-\bbox{\omega})\times \vec{n}]$.  For such
strains to rotate the director in a cholesteric, they would have
to vary along the helical axis. By the requirement of elastic
compatibility this introduces  secondary shears. One can show that
these are prohibitively expensive for this geometry. We
accordingly ignore $\bbox{{\lambda}}$ terms.

\vspace*{-3mm}
\begin{figure}[t!]
  \begin{center}
    \leavevmode
    \epsfxsize=7.5cm
    \epsfbox{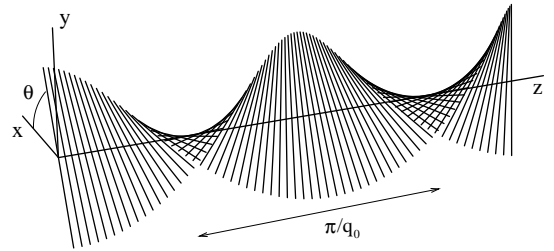}
\caption{A cholesteric director configuration.}
\vspace*{-5mm}
  \label{fig0}
  \end{center}
\end{figure}
Consider an initially regular cholesteric helix along the $z$
axis, Fig.\ 1.  $\theta$ is the azimuthal angle the local director makes
within the $x-y$ plane and is initially
\begin{equation}
\theta_0 (z)= q_0z \label{pitch}
\end{equation}
where $q_0$ is the chiral pitch wavenumber. One can generalise the
$D_1$ term to the large angle limit by considering a molecular
nematic rubber elastic model \cite{mark}.  It is no longer
$\textstyle{1\over 2} D_1 \; (\theta - \theta_0)^2$ but
$\textstyle{1\over 2} D_1 \; \sin^2 (\theta - q_0 z)$ which has
the correct locally nematic symmetry of $\vec{n} \equiv -\vec{n}$.
The energy for an elastomer formed under a cholesteric solvent
which is subsequently replaced with an achiral one is then:
\begin{equation}
F=\int dz\;\; \textstyle{1 \over 2}\; [K_2\; \theta'^2 + D_1 \sin ^2 (\theta - q_0 z)]
\label{basicF}
\end{equation}
where $'$ indicates $d/dz$. The first term in $F$ wishes to remove
the twist in $\vec{n}$ since now the chiral imperative of the
solvent is removed the usual Frank twist penalty is fully
incurred.  However the second term insists that $\vec{n}$ is
anchored to a helix thanks to the crosslinks being formed under
cholesteric conditions. The initial network polymers are taken to
be achiral, thus $q_0$ is only induced and can be tuned with the
choice of the chiral solvent we subject our elastomer to at
formation. If achiral solvent is used when crosslinking, $q_0$
could simply be zero, and we retrieve the description of more
conventional nematic elastomers \cite{mark}.

The two limits of the energy density are (i) the perfectly twisted
cholesteric state $f=\textstyle{1\over 2}K_2 q_0^2$ where the
current pitch wavevector is unchanged from $q_0$, and (ii) the
untwisted state $f=\textstyle{1\over 4}D_1$, the additional factor
of $\textstyle{1\over2}$ arising from the averaging of $\sin^2$
over one period. Thus, crudely, we expect the director to be
twisted if
\begin{equation}
K_2 q_0^2 < D_1/2
\label{condition}
\end{equation}
and untwisted otherwise, this balance being tuneable since $K_2$,
$D_1$ and $q_0^2$ vary relatively to each other with temperature,
degree of crosslinking and swelling and the presence of additional
chiral agents. Eq.\ (\ref{condition}) anticipates the physics of
our detailed results: highly twisted states ($q_0$ large) or
systems with a large twist constant $K_2$ will pay a very high
Frank penalty on loss of spontaneous twist arising from the loss
of the chiral solvent.  A large combination $K_2q_0^2$ will
overcome the anchoring $D_1$ and the elastomer will untwist -
imprinting will be lost.  Weakly twisted elastomers with weak
twist constants will not overcome director anchoring and
imprinting will remain.  We now analyze Eq.\ (\ref{basicF}) for
details of the phase behaviour.

To simplify matters, we begin with the following substitutions: $$
\phi=q_0 z\!-\!\theta \!+\! \pi/2;\quad u=z/\xi; \quad
\xi=\!\sqrt{K_2/D_1}; \quad \alpha=\xi q_0 $$ the angle $\phi$
describes the variation away from (or modulation of) the original
helical pattern, that is we are now in a rotating frame of
reference. Lengths are reduced by the nematic rubber penetration
depth $\xi$, the natural length scale in the problem. The
parameter $\alpha$ is a non-dimensional measure of the nematic
length relative to the chiral pitch, and the condition
(\ref{condition}) is equivalent to $\alpha \sim 1/\sqrt{2}$.
Following the remarks below Eq.\ (\ref{condition}), we expect
imprinting to be lost at large $\alpha$ and retained at small
$\alpha$. Dropping a constant arising from the variable change
$\theta \rightarrow \phi$, the reduced energy (per unit
cross-section area perpendicular to the pitch axis) is:
\begin{equation}
\tilde{F}={2F \over D_1 \xi}=\int du \;\; [(\dot{\phi}-\alpha)^2 - \sin^2 \phi ]
\label{easyF}
\end{equation}
where $(\,\dot{}\,)$ signifies $d/du$. If we make the analogy to
Lagrangian dynamics, the integrand is the Lagrangian density
($L=T-V$) of a particle in a potential given by $\sin^2 \phi$. The
problem also now resembles the problem of an electric or magnetic
field applied perpendicularly to the helix of a liquid
cholesteric,  solved long ago by de Gennes and Meyer \cite{meyer}.
In reduced terms, the $\sin^2\phi$ is like the field
competing with the natural chirality $\alpha$.  In terms of the
original problem, Eq.\ (\ref{basicF}), it is as if a naturally
untwisted nematic has a spatially chiral electric field ``$D_1"$
applied to it in an attempt to induce a twist.

The first integral of the Euler-Lagrange equation corresponding to
Eq.\ (\ref{easyF}) , the elliptic equation or often called in the
literature the Sine-Gordon equation, leads to:
\begin{equation}
\dot{\phi}^2+\sin^2 \phi = c^2
\label{1stintegral}
\end{equation}
where $c^2$ is the integration constant, which has the physical
interpretation of total energy. The analogy is helpful, and before
we examine the details we already foresee two scenarios: If
$c^2<1$, the particle does not have sufficient energy to overcome
the barriers in $\sin^2 \phi$ potential and is therefore
localized, {\it i.e.} $-\pi/2 < \phi < \pi/2$. This corresponds to
the case where cholesteric pitch is largely maintained, with
$\phi(u)$ only introducing small modulations to the director
orientation. If $c^2>1$, our particle can climb out of the
potential valleys and travel freely; this corresponds to the case
of winding/unwinding the cholesteric twists in director
orientation.  Below, we determine $c$ in terms of $\alpha$.


{\it The localised limit, $c^2<1$.} Our particle oscillates
between two values of $\phi_m =\pm \arcsin c$. We accordingly
introduce a new variable $\beta$ in the interval $\beta \in [-{\pi
\over 2}, {\pi \over 2}]$, so that $\sin \phi \equiv c \; \sin
\beta$. Rewriting the derivative $\dot{\phi}$ in terms of
$\dot{\beta}$ and returning it to Eq.\ (\ref{1stintegral}) reduces
this equation to one of the standard elliptic form:
$\dot{\beta}=\sqrt{1-c^2 \sin^2 \beta}$. The period of the
oscillatory motion is found to be:
\begin{equation}
T_1=2 \int_0^{\pi \over 2} \; {d \beta \over \dot{\beta}}= 2\; {\cal K}(c).
\end{equation}
Here ${\cal K}(c)$, and later ${\cal E}(c)$, are the complete
elliptic integrals of the first and second kind respectively. The
period $T_1$ gives, in units of the characteristic length scale
$\xi$, the spatial repeat distance of our $\phi$-modulation of the
original cholesteric angle $q_0 z$. The reduced energy of a period
can be obtained from Eq.\ (\ref{easyF}) as:
\begin{eqnarray}
\tilde{F}_1(T_1)
&=&2[(c^2+\alpha^2-2)\;{\cal K}(c)+\! 2\;{\cal E}(c)].
\end{eqnarray}
In order to compared the stability of various states, we require
the reduced energy {\it density}.  In the localized regime, $c^2 <
1$, we denote this by $g_1(c)$:
\begin{equation}
g_1=\tilde{F}_1(T_1)/T_1=c^2+\alpha^2-2+{2 {\cal E}(c) \over {\cal K}(c)}.
\label{energy1}
\end{equation}
We now require the energy density in the traveling regime.

{\it The traveling limit $c^2>1$.} The modulation period down the
original  helix corresponds to the time taken for our particle to
travel from one peak to the next in the potential.  It is
calculated using Eq.\ (\ref{1stintegral}):
\begin{equation}
T_2=2\int_0^{\pi \over 2} \;{d\phi \over \dot{\phi}}=2k\;{\cal K}(k)
\end{equation}
where $k=1/c$, and thus $k<1$. The reduced energy of a period is:
\begin{eqnarray}
\tilde{F}_2(T_2)
&=&2[(\alpha^2-c^2)k\;{\cal K}(k)+ 2c\;{\cal E}(k)-\alpha \pi].
\end{eqnarray}
The corresponding energy density, $g_2(c)$ for $c^2 >1$, is:
\begin{equation}
g_2=\tilde{F}_2(T_2)/T_2=\alpha^2-c^2+{2 \over k\;{\cal K}(k)}
\left[ {{\cal E}(k) \over k}- {\alpha \pi \over 2} \right].
\label{energy2}
\end{equation}


Combining $g_1(c)$ and $g_2(c)$, Eq.s (\ref{energy1},
\ref{energy2}), we now have the energy density $g(c)$ for the
entire range of $c$. Fig.\ 1 illustrates the energy density plots
for three different values of of the chiral strength, described by
the parameter $\alpha$.

\begin{figure}[t!]
  \begin{center}
    \leavevmode
    \epsfxsize=8cm
    \epsfbox{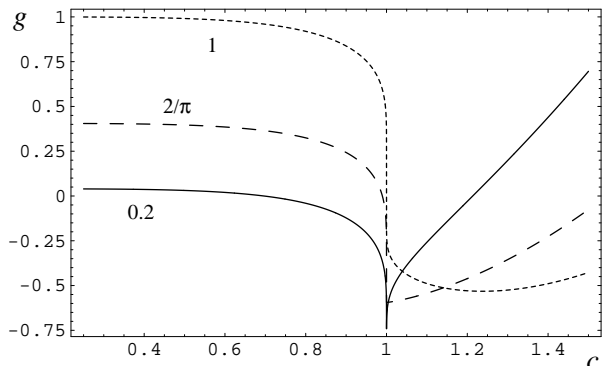}
\caption{Reduced energy density, $g(c)$, for
$\alpha=0.2,\;2/\pi,\;1$.  One has $g_1(c)$ for $c^2 <1$ and
$g_2(c)$ for $c^2 >1$.}
  \label{fig1}
  \end{center}
\end{figure}
We can see that at the transition point $c=1$ the two energy
densities approach each other. However, the densities are not
smoothly joined, forming a cusp which turns from pointing downward
to upward upon the parameter $\alpha$ increasing past a critical
value $\alpha_c=2/\pi$. Invoking the identities for differentials
of complete elliptic integrals: $$ {d {\cal K} \over dk}={{\cal
E}\over k (1-k^2)}-{{\cal K} \over k} \quad ; \; \quad {d {\cal E}
\over dk}={{\cal E}-{\cal K} \over k} $$ we can minimise the
reduced free energy with respect to $k$, or equivalently $c$, by
setting $dg/dc=0$. As evident from Fig.\ \ref{fig1}, minima only
exist in the region $c^2>1$, that is in $g_2(c)$, and only when
$\alpha >\alpha_c$.  There are no minima in $g_1$. The analogous
problem of a cholesteric liquid in the presence of an electric
field \cite{meyer} is similar.  The condition  $dg/dc=0$ fixes $c$
since, for a given $\alpha$, $k$ or $c$ must satisfy:
\begin{equation}
\alpha={2{\cal E}(k) \over \pi k} \quad {\rm for}\quad
\alpha>\alpha_c=2/\pi \; . \label{result}
\end{equation}

The period of $\phi(u)$ modulation is $T=2k \;{\cal K}(k)$ in
units of $\xi$. For each such period along the helical pitch axis,
$\phi$ increases by $\pi$, the director  unwinds once, and
therefore one loses $1$ of the $T\xi q_0/\pi \equiv 2 \alpha k
\;{\cal K}(k)/\pi$ twists imprinted over the interval $T$. The
imprinting efficiency is given by the fractional number of twists
lost:
\begin{equation}
e_0={2\alpha k\;{\cal K}(k) - \pi \over 2\alpha k\; {\cal K}(k)}
\; .
\end{equation}
Provided that $e_0$ is close to unity, much of the chirality can
be preserved.

For small chiral power $\alpha<\alpha_c = 2/\pi$, the minimising
condition $dg/dc=0$ has no solution and the minimum free energy
occurs exactly at $c=1$. The cusp has the energy density
$g=(2/\pi)^2-1$, and a logarithmically divergent period $T$ which
implies $e_0=1$. The director gets arrested, attempting to untwist
but never actually manages a full turn within a finite distance.
The imprinting is therefore successful.

For chiral power $\alpha > \alpha_c = 2/\pi$, minima are found for
$c>1$.  The period is no longer infinite and twists are lost. For
large $\alpha$, {\it i.e.} the nematic penetration depth large
compared to the cholesteric pitch, we can expand the elliptic
function for small $k$ to find $\alpha\sim c(1-k^2/4-3k^4/128 ...
)$. Alternatively $k\sim 1/\alpha(1-1/4\alpha^2 ...)$ and we find
a period of $T=2k\;{\cal K}(k) \sim \pi/\alpha$ in units of $\xi$.
Thus for every actual distance of $\pi \xi/ \alpha=\pi/ q_0$,
$\phi$ accumulates an increment of $\pi$. That is, $e_0\rightarrow
0$,  corresponding eventually to the case of complete unwinding of
the imprinted helical pattern.

%
\begin{figure}[t!]
  \begin{center}
    \leavevmode
    \epsfxsize=6cm
    \epsfbox{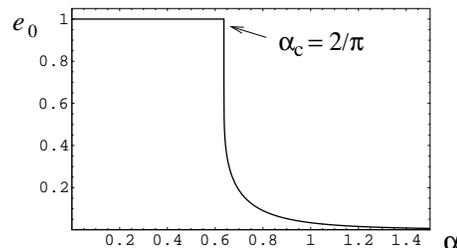}
    \caption{Imprinting efficiency vs.\ network parameter $\alpha$. The slope
at the critical point $\alpha=\alpha_c^+$ diverges and the efficiency
decays rapidly.}
  \label{fig2}
  \end{center}
\end{figure}
%

In writing equation (\ref{basicF}), we have assumed that the
chiral solvent was completely replaced with an achiral one as was
the case in experiment \cite{geof}. However, we can trivially
generalise $F$ to the case where the current solvent is chiral:
\begin{equation}
F=\int dz\;\; \textstyle{1 \over 2}\; [K_2\; (\theta'-q)^2 + D_1
\sin ^2 (\theta - q_0 z)] \; ,\label{gerneralF}
\end{equation}
where $q$ is the wave number that the solvent would introduce in
the absence of crosslinks.
Our previous case corresponds to $q=0$. The mathematical procedure
remains applicable, with an adjustment of the parameter
$\alpha=\xi q_0$ to: $$ \alpha'=\xi (q_0 - q) $$ which indicates a
chiral environment of the same handedness can help preserve the
chirality by reducing the value of $\alpha$ (negative values of
$\alpha$ are mathematically equivalent to $-\alpha$,
if we switch the sign of the $\phi$ modulation). A chiral environment
of the other handedness, $q<0$, instead increases $\alpha$ and has
the opposite effect.  It reduces imprinting, as one might expect.
By simply varying $q$, via the solvent composition, we have a
powerful way to map out the wind-unwinding transition.

Another interesting case is the induction of a cholesteric state
in an initially uniform nematic network upon introducing 
a chiral solvent \cite{initexp}. That is $q_0=0$ and $\alpha'=\xi q$,
where $q$ is the chiral pitch wave number due to the chiral solvent.
In this case, $e_0$ given in Fig.\ 2 can be interpreted as the 
network resistance to imprinting - until the point $\alpha'$
reaches $2/\pi$ there will be no helicity induced in the network
at all, and thereafter it rises with $\alpha'$.

Finally, by varying the amount of solvent in the crossed network, we can also tune, {\it i.e.}
contract or expand, the volume relative to its original: $V \rightarrow \beta V$.
With this, we expect from molecular interpretations,
$K_2\rightarrow K_2 $ for a special case where the local nematic order is preserved
(the nematic order variation due to the solvent or temperature will be examined elsewhere),
but $D_1 \rightarrow D_1/\beta$ as it depends on
crosslink density, and finally assuming an isotropic expansion/contraction,
$q_0 \rightarrow q_0 \beta^{-1/3}$. Thus we have
$$\alpha \rightarrow \alpha \beta^{1/6},$$
so swelling can, albeit weakly, increase the parameter $\alpha$ in our model and
discourage the preservation of the imprinted chirality.

In conclusion,
we have proposed a continuum model for chiral imprinting in nematic
elastomers. The model predicts the residual chirality when the chiral
solvent is removed. A twist-untwist transition emerges from the
theory, and an experimental verification of this transition
would be a useful test of the theory presented here. Further
work is focussed on the  mechanical properties of networks; we expect
an external mechanical field will have substantial influence on the
imprinting of chirality.

\vspace{0.2in} We thank E.~M.~Terentjev, R.~B.\ Meyer,  and M.~E.\
Cates for useful discussions.  The problem was first suggested to
us by H. Finkelmann. YM is grateful to St John's College,
Cambridge for a research fellowship.


\end{multicols}\end{document}